\documentclass[letter]{aa} 

\usepackage{graphicx, comment}
\usepackage{commath}
\usepackage{txfonts}
\usepackage{subfig}
\usepackage{upgreek}
\usepackage[colorlinks=true]{hyperref}

\hypersetup{
     colorlinks   = true,
     citecolor    = blue
}

\usepackage{orcidlink}
\newcommand{\F}{Fornax\,A}

\newcommand{\dlim}{70\,\upmu\textrm{as}}

\newcommand{\Tbmax}{1.7\times10^9\,\textrm{K}}
 
\newcommand{\TbcleanRzero}{(3.4\pm0.3)\times10^{11}\,\textrm{K}}
\newcommand{\Tblim}{(1.5\pm0.2)\times10^{11}\,\textrm{K}} 
\newcommand{\Tbehtim}{(8.1\pm0.8)\times10^{12}\,\textrm{K}}

\newcommand{\SzeroCLEAN}{62.3\pm6.2\,\textrm{mJy}}
\newcommand{\RCLEAN}{96.0\,\upmu\textrm{as}}

\newcommand{\SzeroEHTIM}{47.5\pm4.8\,\textrm{mJy}} 
\newcommand{\REHTIM}{10.0\,\upmu\textrm{as}}

\newcommand{\SzeroRange}{47.5-62.3\,\textrm{mJy}}

\newcommand{\cpscale}{4000\,\upmu\textrm{as}} 
\newcommand{\conf}{91.7\%}
\newcommand{\detec}{8.3\%}

\begin{document}

    \title{First VLBI detection of \F}

   \author{
   G.~F. Paraschos\inst{1} \orcidlink{0000-0001-6757-3098},
   M. Wielgus\inst{1,2} \orcidlink{0000-0002-8635-4242},
   P. Benke\inst{1}\orcidlink{0009-0006-4186-9978}, 
   V. Mpisketzis\inst{3, 4}\orcidlink{0009-0009-9682-3119}, 
   F. Rösch\inst{5}\orcidlink{0009-0000-4620-2458},
   K. Dasyra\inst{3}\orcidlink{0000-0002-1482-2203},
   E. Ros\inst{1}\orcidlink{0000-0001-9503-4892},
   M.~Kadler\inst{5}\orcidlink{0000-0001-5606-6154},
   R. Ojha\inst{6}, 
   P. G. Edwards\inst{7}\orcidlink{0000-0002-8186-4753}, 
   L. Hyland\inst{8}\orcidlink{0000-0002-4783-6679},
   J. F. H. Quick\inst{9},
   S. Weston\inst{10}
          }

   \authorrunning{G.~F. Paraschos et al.}
   \institute{
              $^{1}$Max-Planck-Institut f\"ur Radioastronomie, Auf dem H\"ugel 69, D-53121 Bonn, Germany\\ 
              $^{}$\ \email{gfparaschos@mpifr-bonn.mpg.de}\\
              $^2$Institute of Physics, Silesian University in Opava, Bezru\v{c}ovo n\'{a}m. 13, CZ-746 01 Opava, Czech Republic\\
              $^3$Department of Physics, National and Kapodistrian University of Athens, Panepistimiopolis, 15783 Zografos, Greece\\
              $^4$Institut für Theoretische Physik, Goethe Universität Frankfurt, Max-von-Laue-Str.1, 60438 Frankfurt am Main, Germany\\
              $^5$Julius Maximilians University Würzburg, Faculty of Physics and Astronomy, Institute for Theoretical Physics and Astrophysics, Chair of Astronomy, Emil-Fischer-Str. 31, 97074 Würzburg, Germany\\
              $^6$NASA HQ, 300 E St SW, Washington, DC 20546-0002, USA\\
              $^7$CSIRO Space and Astronomy, PO Box 76, Epping, NSW1710, Australia\\
              $^8$School of Natural Sciences, University of Tasmania, Private Bag 37, Hobart, Tasmania 7001, Australia\\
              $^9$Hartebeesthoek Radio Astronomy Observatory, PO Box 443, 1740 Krugersdorp, South Africa\\
              $^{10}$Space Operations New Zealand Ltd, 62 Deveron Street, Invercargill 984, New Zealand
             }

   \date{Received -; accepted -}

 \abstract
   {
Radio galaxies harbouring jetted active galactic nuclei are a frequent target of very-long-baseline interferometry (VLBI) because they play an essential role in exploring how jets form and propagate.
Hence, only few have not been detected with VLBI yet; \F\ is one of the most famous examples.
Here we present the first detection of the compact core region of \F\ with VLBI.
At 8.4\,GHz the faint core is consistent with an unresolved point source. 
We constrained its flux density to be $S_0=\SzeroRange$ and its diameter to be $D_0^\textrm{min} \leq \dlim$.
The high values of the measured brightness temperature ($T_\textrm{B}\gtrsim10^{11}\,\textrm{K}$) imply that the observed radiation is of non-thermal origin, likely associated with the synchrotron emission from the active galactic nucleus.
We also investigated the possibility of a second radio source being present within the field of view.
Adding a second Gaussian component to the geometrical model-fit does not significantly improve the quality of the fit and we, therefore, conclude that our detection corresponds to the compact core of \F.
Analysis of the non-trivial closure phases provides evidence for the detection of more extended flux density, on the angular scale of $\sim\cpscale$.
Finally, the fractional circular polarisation of the core is consistent with zero, with a conservative upper limit being $m_\textrm{circ}\leq4\%$.
   }


   \maketitle

\section{Introduction}

Relativistic astrophysical jets emanating from the centres of galaxies are a telltale sign of highly energetic activity connected to the central supermassive black hole (SMBH) that they harbour.
Specifically, radio galaxies with an active galactic nucleus (AGN) are a ubiquitous target of jet studies, since they are generally nearby and therefore provide more insights into jet formation and launching.
At larger scales, jet interactions with the ambient medium can give clues about AGN feedback, star formation, and galactic evolution \citep[e.g.][]{Fabian00, Burillo14, Fotopoulou19, Dasyra22}.
On the other hand, at smaller scales, the ultimate vicinity of SMBHs can be explored using very-long-baseline interferometry (VLBI), with which the shrouding medium around them can be peered through.
Such studies have been carried out, for example, for M\,87 \citep{EHT19a, Lu23}, Sgr\,A$^{*}$ \citep{EHT22a}, Cygnus\,A \citep{Boccardi16}, 3C\,84 \citep{Nagai14, Paraschos21, Paraschos22}, NGC\,1052 \citep{Baczko16}, and Centaurus\,A \citep{Mueller11, Mueller14, KimJH18, Janssen21} among others.

\F\ (NGC\,1316, J0322-3712) is another such radio galaxy whose radio emission is very prominent, providing clues about its central engine.
It is located in the Fornax cluster and is one of the nearest \citep[$D_\textrm{L} = 18.8\,\textrm{Mpc}$;][]{Hatt18} and brightest \citep[e.g.][]{Cantiello13} radio galaxies, harbouring a SMBH of $\sim1.3\times10^8\,\textrm{M}_\odot$ \citep{Nowak08}.
This system is thought to be the result of more than one merger event \citep{Schweizer80, Mackie98, Lanz10, Richtler12a, Richtler12b}.
Consequently, it has been the target of numerous studies, covering a wide frequency spectrum \citep[see][among others]{Stanley50, Schweizer80, Geldzahler84, Horellou01, Lanz10, Richtler12a, Maccagni20}.
The most prominent morphological characteristic of \F\ are two extended radio jets \citep[the centres of the two lobes are $\sim30$\,arcmin apart, with total edge-to-edge size $\leq50$\,arcmin; see e.g.][]{Ekers83} emanating from the central AGN, encased in dense interstellar medium.
Their larger scale physical characteristic have been studied with radio and X-ray observations, revealing variability in the orientation of the inner jets \citep{Geldzahler84} and associated X-ray cavities \citep{Lanz10}.
Most recently, MeerKat data showed that the radio lobes of \F\ have a double shell morphology and that its core is rapidly flickering \citep{Maccagni20}.
The complexity of its structure has been connected to past merger events \citep{Lanz10, Richtler12a, Richtler12b}, possibly associated with its numerous occurrences of intermittent radio activity \citep{Maccagni20}.

Nevertheless, \F\ has not been detected with VLBI thus far \citep{Slee90, Fomalont00, Angioni19}.
This is because VLBI monitoring of the southern sky is limited and this source exhibits a faint nuclear region.
With our exploratory observations, we aimed to study the core region of \F, peering as far as possible into the vicinity of the central engine.
Here we report the first VLBI detection of this compact region of \F, with which we were able to constrain the size and structure of its core.
This work is structured as follows: In Sect.~\ref{sec:Results} we present the observational setup, our data analysis methodology, and our results. 
In Sect.~\ref{sec:Discussion} we discuss constraints of the core structure and polarisation and in Sect.~\ref{sec:Conclusions} we conclude with a summary of this work.

\section{Observations, data analysis, and results}\label{sec:Results}

\begin{figure}
\centering
\includegraphics[scale=0.3]{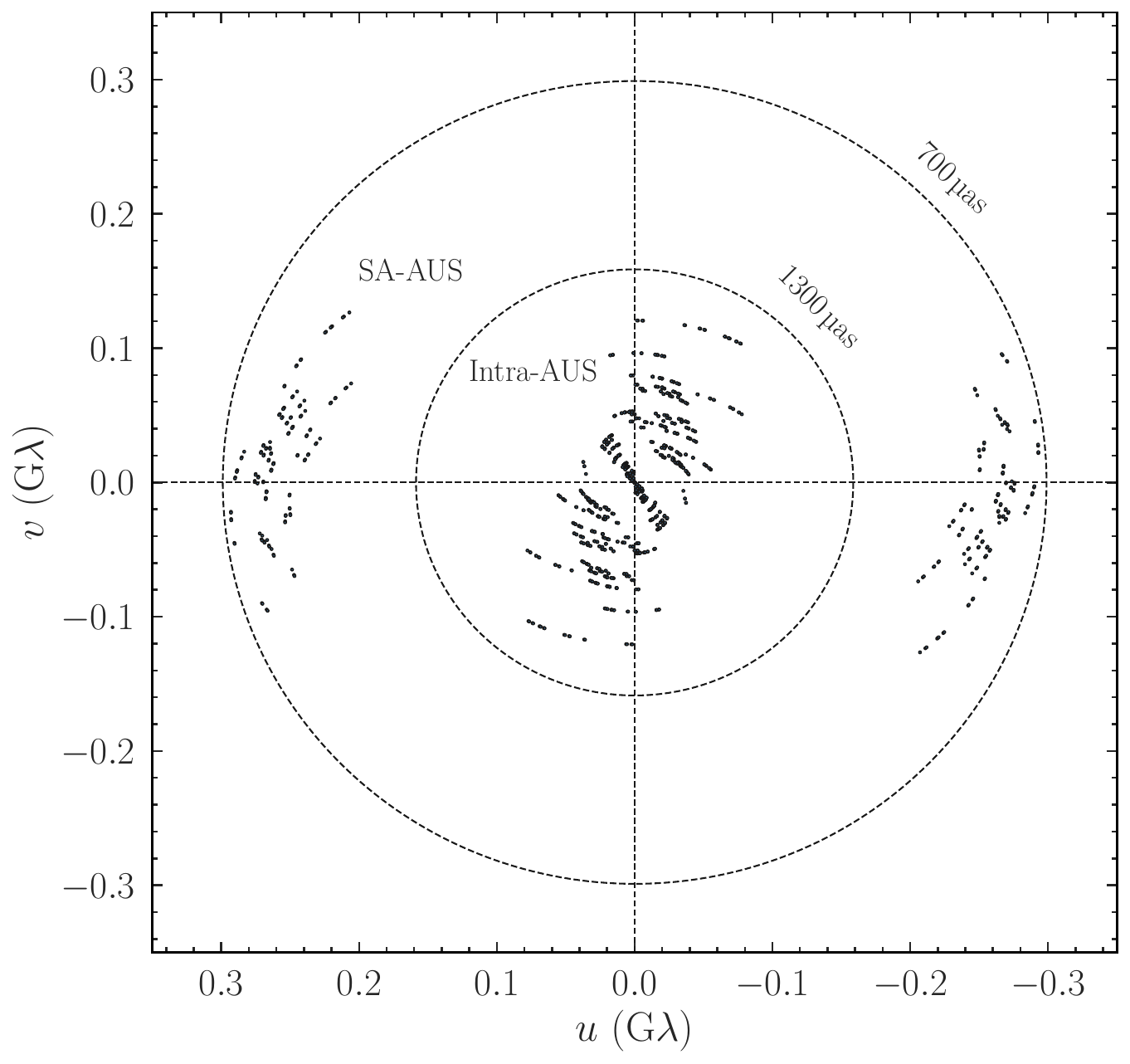}
  \caption{
  Illustration of the TANAMI $(u, v)$-coverage of the target source \F.
  Two point clouds can be identified in the $(u, v)$-coverage, corresponding to the shorter distance Intra-Australian/New Zealand baselines and to the ones between Australia/New Zealand and South Africa.
  The outer dashed circle corresponds to a $(u,v)$-radius of 700\,micro-arcseconds ($\upmu$as) and the inner one to 1300\,$\upmu$as.
  }
     \label{fig:UV}
\end{figure}

\F\ was observed in $X$-band (8.4\,GHz) on February 11 2023 for a total of one hour as a test source, as part of the VLBI monitoring program Tracking Active Galactic Nuclei with Austral Milliarcsecond Interferometry \citep[TANAMI; see][]{Ojha10}.
The participating antennas ATCA (AT; $5\times22\,\textrm{m}$), Ceduna (CD; $30\,\textrm{m}$), Hobart (HO; $26\,\textrm{m}$), Katherine (KE; $12\,\textrm{m}$), Mopra (MP; $22\,\textrm{m}$), Parkes (PA; $64\,\textrm{m}$), DSS36 (TD; $34\,\textrm{m}$), and DSS43 (TI; $70\,\textrm{m}$) in Australia, Warkworth (WW; $12\,\textrm{m}$) in New Zealand, and Hartebeesthoek (HH; $26\,\textrm{m}$) in South Africa.
Further details about them can be found in \cite{Benke24}.
In Fig.~\ref{fig:UV} we display the $(u, v)$-coverage corresponding to observations of \F\ with the TANAMI array.

We conducted the data calibration with the software package \texttt{AIPS} \citep{Greisen03}.
The standard VLBI procedure was followed \citep[described in more detail in][]{Benke24}, in which digital sampling and parallactic angle corrections were applied to the data and fringe fitting and amplitude calibration were performed.
A fringe fitting signal-to-noise ratio ($S/N$) detection threshold of $5\,\sigma$ was imposed.
Phase-referenced calibration was also performed, however, this approach did not yield an increase in the number of detections and the final data products did not exhibit any discernible difference.
Finally, the data were averaged in frequency (to output a single channel per sub-band) and coherently in time (into ninety second bins), and then written out for image reconstruction. 
The final $S/N$ in the range of $S/N\sim5-30$ provided a solid basis for our subsequent analysis.

\begin{figure*}
\centering
\includegraphics[scale=0.35]{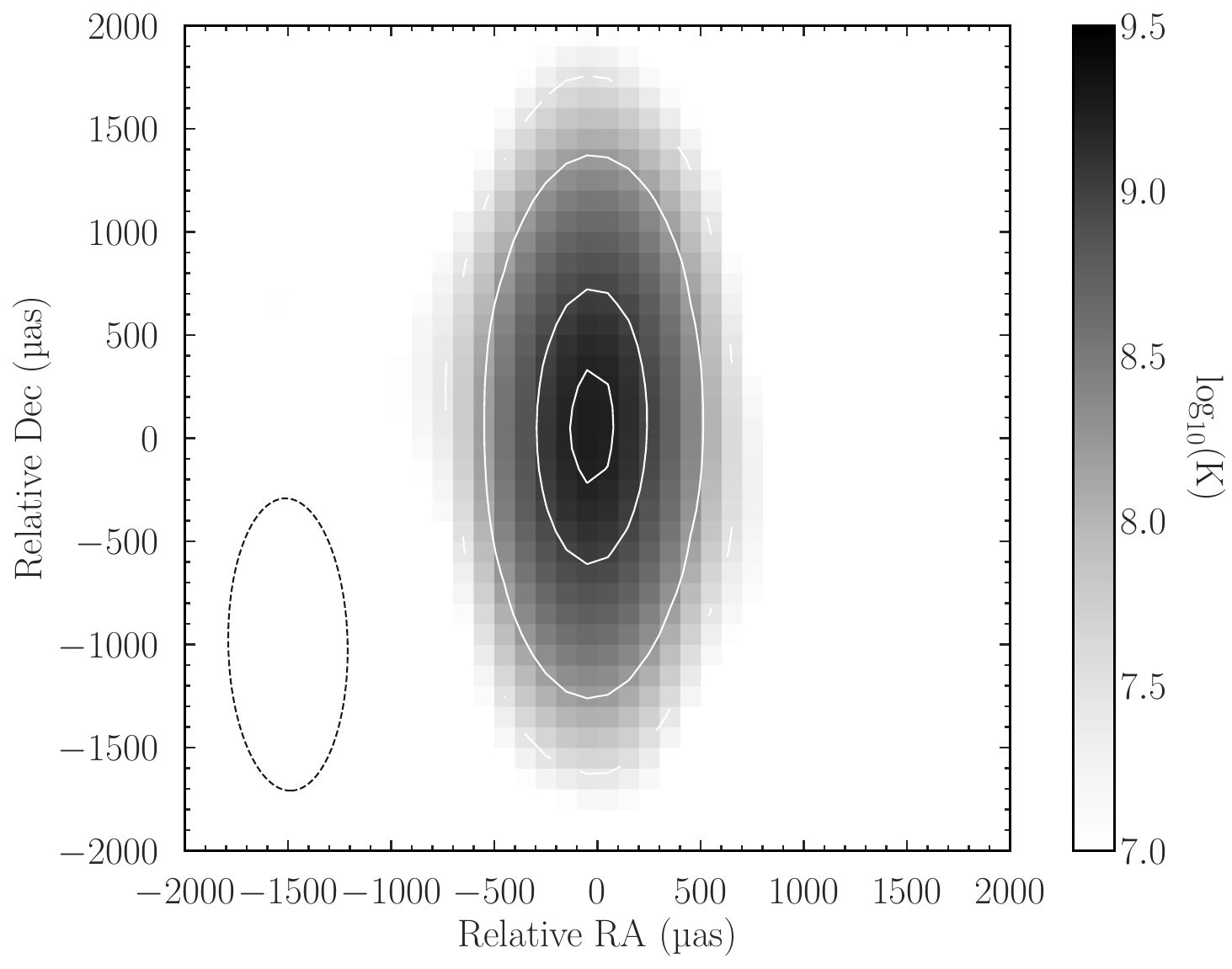}
\includegraphics[scale=0.342]{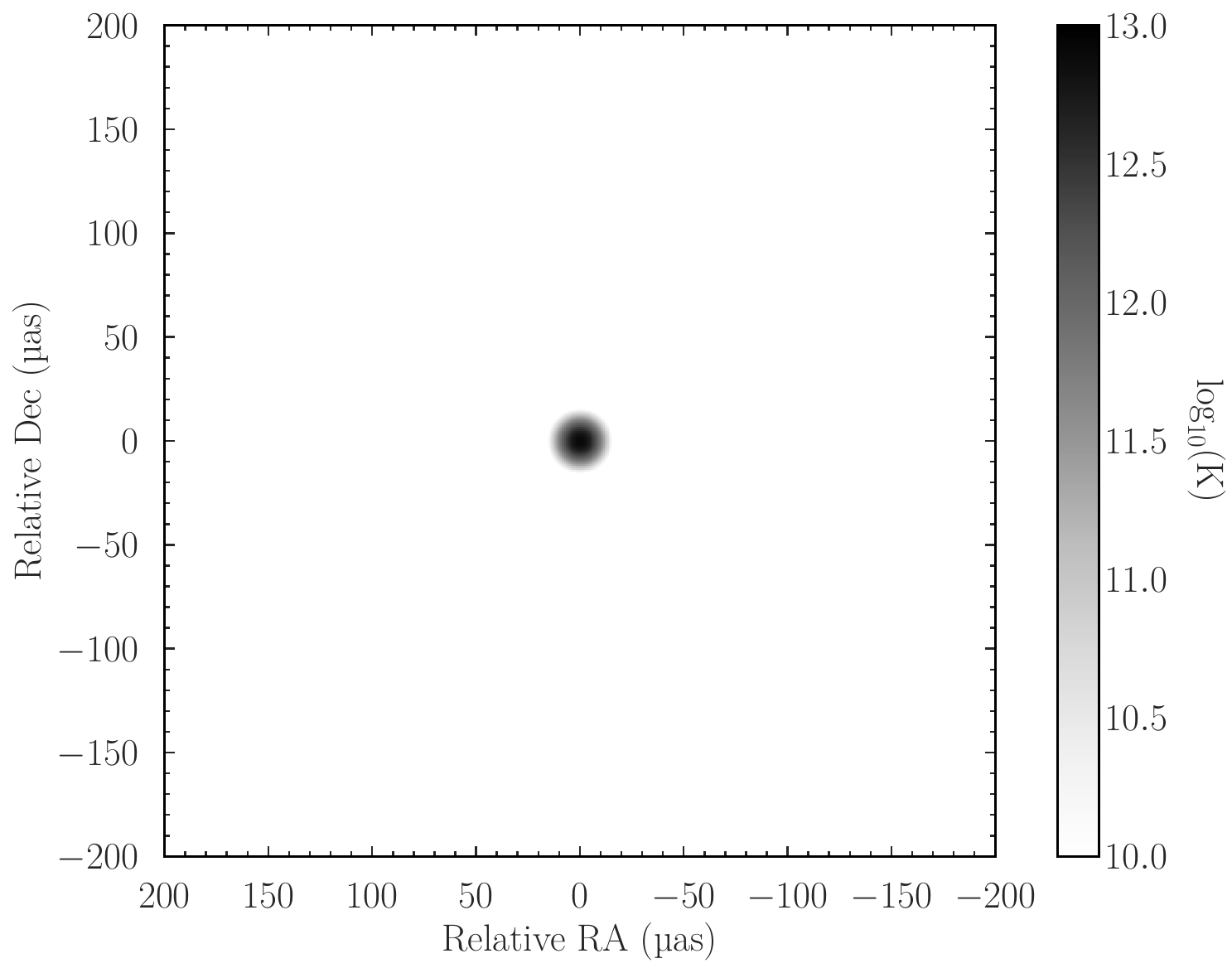}
\includegraphics[scale=0.35]{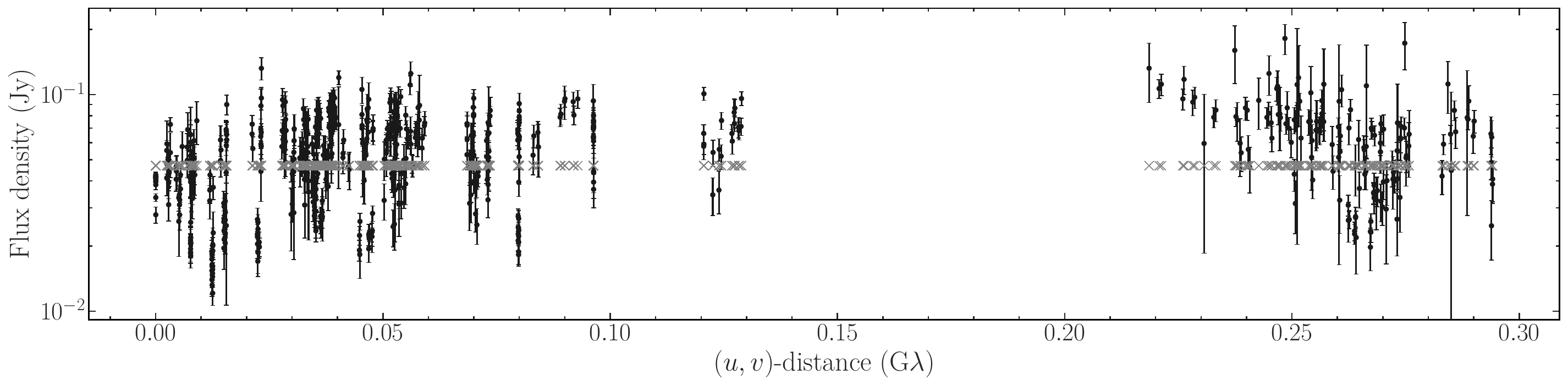}
  \caption{
    Geometrical model of \F\ using \texttt{CLEAN} and geometrical modelling image reconstruction and flux density versus $(u, v)$-distance plot.
    \textit{Left}: \texttt{CLEAN} image of \F.
    The structure is consistent with that of a point source, convolved with an elliptical beam of size $580\times1420\,\upmu\textrm{as}$ (indicated with a dashed black ellipse in the bottom left corner).
    The contour levels are 1, 5, 30, 50, 75, and 95\% of the intensity maximum of $\Tbmax$ and the colour scale corresponds to the measured $T_\textrm{B}$.
    \textit{Right}: \texttt{eht-imaging} model of \F. 
    The point source structure is confirmed.
    The higher values recovered here stem from the fact that, in the geometrical modelling framework, beam convolution is not required and thus the small diameter of the Gaussian component plugged into Eq.~\ref{eq:Tb} increases the corresponding $T_\textrm{B}$.
    Notice here that the axes are zoomed in by a factor of ten compared to the left panel.
    \emph{Bottom}: The data and their uncertainties are shown with black dots and the best-fit model is overlaid with grey crosses.
    The preferred flux density level of the single Gaussian component of the model at $S^\textrm{ehtim}_0 = \SzeroEHTIM$ is clearly visible.
  }
     \label{fig:StokesI}
\end{figure*}

The challenges of the analysis of the \F\ data set are caused by the sparse $(u, v)$-coverage with a large gap between approximately 0.1 and 0.2\,G$\uplambda$ (see Fig.~\ref{fig:UV}) related to the geographic spread of the antennas that comprise TANAMI and uncertainties of the amplitude gain calibration in the absence of strongly constraining interferometric closure information. 
To leverage relatively high $S/N$ and to mitigate potential spurious instrumental effects or artefacts in our data, we employed two methods for the source morphology reconstruction.
Our first approach was to use the \texttt{CLEAN} algorithm within the \texttt{difmap} software package \citep{Shepherd94}.
The image is shown in the left panel of Fig.~\ref{fig:StokesI}; the structure is consistent with a compact point source.
As a second step we modelled the flux density with a circular Gaussian component. 
The best fit parameters for flux density ($S_0^\textrm{clean}$) and diameter ($D_0^\textrm{clean}$) were: $S_0^\textrm{clean} = \SzeroCLEAN$ and $D_0^\textrm{clean} = \RCLEAN$.

Our second approach was to use geometrical modelling within the \texttt{eht-imaging} framework \citep{Chael16, Roelofs23}.
The advantage of this forward-modelling technique lies in the fact that best-fit solutions are leveraged from the data without the necessity of convolving the solution afterwards with the point-spread function.
Thus, more fine structure can be recovered and information can be extracted even from sparse data-sets \citep[see e.g.][]{Paraschos24}, without performing full-fledged imaging.
The right panel of Fig.~\ref{fig:StokesI} presents our best-fit model to the visibility amplitudes and closure phases as a function of the $(u, v)$-distance. 
In this case the input model used was an elliptical Gaussian component.
However, the best fit was achieved when the major axis of the ellipse equalled the minor one, resulting again in a circular Gaussian component, characterised by a flux density of $S_0^\textrm{ehtim} = \SzeroEHTIM$ and a diameter of $D_0^\textrm{ehtim} = \REHTIM$.
We note here that $D_0^\textrm{ehtim}$ corresponds to the lower limit of the allowed parameter extent.

Using the formalism described in \cite{Lobanov05}, we then calculated the minimum resolvable scale of our observations, which corresponds to $D_\textrm{lim} = 70\,\upmu\textrm{as}$.
Since from our two approaches we constrained diameter of the compact source to a lower value on average, this implies that the compact source is consistent with an unresolved point source ($D_0^\textrm{min} \leq \dlim$).
The bottom panel of Fig.~\ref{fig:StokesI} showcases the fitted data and the geometrical model used.
The measurements at longer baselines correspond to robust detections, with an $S/N$ of $\geq5$, indicative of a very compact structure.

We also constrained the flux density of the source's core to be $S_0 = \SzeroRange$.
In order to transform this into the observed brightness temperature $T_\textrm{B}$, we used the formula:
\begin{equation}
    T_\textrm{B} = 1.22\times10^{15}\left(\frac{S}{\textrm{mJy}}\right)\left(\frac{\nu}{\textrm{GHz}}\right)^{-2}\left(\frac{\phi}{\upmu\textrm{as}}\right)^{-2}, \label{eq:Tb}
\end{equation}
where $\nu$ is the observing frequency and $\phi$ is the source size (or a corresponding upper limit for an unresolved source).
For the \texttt{CLEAN} reconstruction (using $\phi=D_0^\textrm{clean}$) this corresponds to $T_\textrm{B} = \TbcleanRzero$.
A more realistic value can be obtained instead by plugging in $\phi=D_\textrm{lim}$, which provides a robust lower limit of $T_\textrm{B} = \Tblim$.
Finally, for the geometrical modelling reconstruction, where no convolution takes place, setting $\phi=D_0^\textrm{ehtim}$ yields $T_\textrm{B} = \Tbehtim$; this result indicates that even higher $T_\textrm{B}$ are feasible for \F.
Overall, the magnitude of $T_\textrm{B}$ suggests that the radiation is non-thermal, associated with synchrotron processes and thus likely connected to its AGN launching the jets of \F.

\section{Discussion}\label{sec:Discussion}

\subsection{Core structure}\label{ssec:CoreStructure}

Given the complexity of the large scale radio lobes connected to the jet emission, it is reasonable to assume that the compact region could be fitted with a core-jet structure. 
In order to test this, we added a second Gaussian component to the geometrical model fit and allowed its position to vary within the entire field of view.
This approach serves two purposes; firstly, it provides a constraint on the core structure.
Secondly, as is the case with wide field imaging, it ensures that no other source is present in the data, thus confirming that our detection corresponds to the radio core of \F.
We found no significant improvement to our fiducial model of one elliptical Gaussian component.
The second one is also preferably placed at the centre of the image and its corresponding flux density tends to the minimum prior value.

Finally, we examined the structure encoded in the closure phases; any discernible structure present in them is suggestive of the existence of asymmetry in the source morphology.
To quantify this, we examined the mean deviation of the closure phases per triangle from zero degrees.
We found that the triangle formed by CD-TD-WW exhibits a deviation of $3.1\,\sigma$ from zero degrees, by time-averaging the closure phases under a constant value assumption. 
In order to assess the significance of this detection, we performed a simple Monte Carlo experiment, modelling closure phases on all independent triangles as normal random variables with zero mean and standard deviation of one. 
We then asked how often the largest measured deviation from zero is larger than $3.1\,\sigma$, obtaining a frequency of $\detec$ as a result. 
Thus, we conclude with $\conf$ confidence, that the CD-TD-WW triangle closure phase deviates from zero, and hence a faint source structure on scales $\sim\cpscale$, probed by that triangle, is present.

\subsection{Polarisation}\label{ssec:Polarisation}

The observations discussed in this work were conducted in full Stokes.
However, extracting polarisation information is challenging due to the faintness of the source, the absence of suitable calibrators, and due to the heterogeneous nature of the array \citep[but demonstrated to be possible, for example, in][]{Dodson08}.
Nevertheless, a statement can be made about the presence of circular polarisation in the signal.
We imaged both parallel hand polarisations in order to investigate whether or not we could detect a mismatch in their respective flux densities.
A significant discrepancy would indicate the detection of a circular polarisation signature in the data \citep{EHT23}.
We found that the difference in flux density between the two parallel hands is of the order of $\sim4\%$.
We note, however, that this value corresponds to a very conservative upper limit, given that the $S/N$ is not significant enough to draw a robust conclusion.
Therefore, the fractional circular polarisation ($m_\textrm{circ}$) of the core of \F\ is consistent with zero, at most $m_\textrm{circ}\leq4\%$.

\section{Conclusions} \label{sec:Conclusions}

In this work we presented the first detection of the radio core of \F\ with VLBI.
Our findings can be summarised as follows:
\begin{itemize}
    \item The core of \F\ is best approximated by an unresolved point source. 
    \item The best-fit model is that of a circular Gaussian component with an upper limit size of $D_0^\textrm{min} \leq \dlim$ and a flux density of $S_0 = \SzeroRange$. 
    \item We constrained $T_\textrm{B}\gtrsim10^{11}\,\textrm{K}$, which is suggestive of non-thermal origin for the measured radiation, possibly originating from the central SMBH of \F.
    \item While through geometrical modelling no additional flux density was recovered, information from closure phases implies the existence of a more extended structure, the size of $\sim\cpscale$, at a confidence level of \conf. 
    \item Fractional circular polarisation of the core region is consistent with zero. 
    A conservative upper limit is of the order of $m_\textrm{circ}\leq4\%$.
\end{itemize}

Our first successful detection of \F\ serves as a proof of concept for the detectability of this source with VLBI.
Future VLBI observations including more sensitive arrays like the next generation Very Large Array, Square Kilometre Array and the Atacama Large Millimeter Array will provide further insight into the compact region of \F. 

\begin{acknowledgements}
      
      This research is supported by the European Research Council advanced grant “M2FINDERS - Mapping Magnetic Fields with INterferometry Down to Event hoRizon Scales” (Grant No. 101018682). 
      The Long Baseline Array is part of the Australia Telescope National Facility (\url{https://ror.org/05qajvd42}) which is funded by the Australian Government for operation as a National Facility managed by CSIRO. 
      From the 2023 July 1, operation of Warkworth was transferred from Auckland University of Technology (AUT) to Space Operations New Zealand Ltd.
      Hartebeesthoek is a facility of the National Research Foundation, South Africa. 
      This research has made use of the NASA/IPAC Extragalactic Database (NED), which is operated by the Jet Propulsion Laboratory, California Institute of Technology, under contract with the National Aeronautics and Space Administration. 
      This research has also made use of NASA's Astrophysics Data System Bibliographic Services. 
      
      Finally, this research made use of the following python packages: {\it numpy} \citep{Harris20}, {\it scipy} \citep{2020SciPy-NMeth}, {\it matplotlib} \citep{Hunter07}, {\it astropy} \citep{2013A&A...558A..33A, 2018AJ....156..123A} and {\it Uncertainties: a Python package for calculations with uncertainties.
      }
\end{acknowledgements}

\bibliographystyle{aa}
\bibliography{aanda}

\end{document}